\documentclass[amstex,aps,preprint,superscriptaddress]{revtex4}%
\usepackage{amsmath}
\usepackage{amsfonts}
\usepackage{amssymb}
\usepackage{graphicx}%
\usepackage{dsfont}

\begin{document}
	\title{Unambiguous joint detection of spatially separated properties of a
	single photon in the two arms of an interferometer}
\author{Surya Narayan Sahoo}
\affiliation{Light and Matter Physics, Raman Research Institute, Bengaluru 560080, India}
\author{Sanchari Chakraborti}
\affiliation{Light and Matter Physics, Raman Research Institute, Bengaluru 560080, India}
\author{Som Kanjilal}
\affiliation{Center for Astroparticle Physics and Space Science (CAPSS), Bose Institute,
Kolkata 700 091, India}
\author{Dipankar Home}
\affiliation{Center for Astroparticle Physics and Space Science (CAPSS), Bose Institute,
Kolkata 700 091, India}
\author{Alex Matzkin}
\affiliation{Laboratoire de Physique Th\'eorique et Mod\'elisation, CNRS Unit\'e 8089, CY
Cergy Paris Universit\'e, 95302 Cergy-Pontoise cedex, France}
\author{Urbasi Sinha}
\email{usinha@rri.res.in}
\affiliation{Light and Matter Physics, Raman Research Institute, Bengaluru 560080, India}

\begin{abstract}
The quantum superposition principle implies that a particle entering an interferometer evolves by simultaneously taking both arms. If a non-destructive, minimally-disturbing interaction coupling a particle property to a pointer is implemented on each arm while maintaining the path superposition, quantum theory predicts that, for a fixed state measured at the output port, certain particle properties can be associated with only one or the other path. Here we report realization of this prediction through joint observation of the spatial and polarization degrees of freedom of a single photon in the two arms of an interferometer. Significant pointer shifts ($\sim$50  microns) are observed in each arm. This observation, involving coupling distinct properties of a quantum system in spatially separated regions, opens new possibilities for quantum information protocols and for tests of quantumness for mesoscopic systems.
	
\end{abstract}
\maketitle


\bigskip

\bigskip
Manifestations of the quantum superposition of states in a two-arm
interferometer open avenues for empirically probing intriguing questions, such
as whether it is possible to jointly detect signatures of distinct particle
properties in different arms of an interferometer.
Indeed, a quantum particle, say a single photon entering an interferometer is said to
travel along both arms simultaneously (1). This is generally evidenced by
monitoring the resulting interference at the exit port. Instead, if a
measurement is made earlier on one or the other arm, the photon will be
detected on that arm with some probability and the interference pattern will
disappear. Modifying an interaction at an intermediate time, such as removing
the exit beam-splitter once the photon is already inside the interferometer in
the famous delayed-choice experiment (2) changes the observed properties of
the photon. More generally, an intermediate non-destructive measurement of a
given observable can be made on one of the arms and the photon is then
measured at the exit port and filtered to a chosen final state (3). The photon
will then be found to have a given value of the measured property on that arm,
with a probability conditioned by the final state (as confirmed by a recent
experiment (4)). For certain observables like projectors giving a yes/no
answer, an intermediate outcome ``yes'' can be obtained with a unit
conditional probability on a given arm, say arm A (and consequently zero
probability on the other arm B). Therefore the particle will always be found
with the property associated with that observable on arm A, never on the
other. However, if a different observable is measured, the particle will
sometimes be found (in an eigenstate of the measured observable) on arm B.

\ \ Consider a single photon entering the Mach-Zehnder interferometer (MZI) of
Figure\ 1. We prepare the state after the beam-splitter BS1 to be \ $\left\vert
\psi\right\rangle =\frac{1}{\sqrt{2}}\left(  \left\vert A\right\rangle
\left\vert H\right\rangle +\left\vert B\right\rangle \left\vert V\right\rangle
\right)  $, where $\left\vert A\right\rangle $ \ and $\left\vert
B\right\rangle $ \ denote the spatial wavefunctions in arms A and B
respectively; this preparation procedure (2) is known as \textquotedblleft
pre-selection\textquotedblright. When detecting the photon at the exit port,
we filter the measured state keeping only the outcomes corresponding to
$\left\vert \phi\right\rangle =\frac{1}{\sqrt{2}}\left(  \left\vert
A\right\rangle +\left\vert B\right\rangle \right)  \left\vert H\right\rangle
$. This filtering procedure is known as \textquotedblleft
post-selection\textquotedblright. Let $\widehat{Y}_{i}=\left\vert
i\right\rangle \left\langle i\right\vert {\otimes}\widehat{\mathds{1}}$
\ \ represent the spatial projection operator on arm $i=$A, B and let
$\widehat{X}_{i}=\left\vert i\right\rangle \left\langle i\right\vert {\otimes
}\widehat{\sigma}_{1}$ \ represent a diagonal polarization measurement on arm
$i$ ($\sigma_{1}$ is a Pauli matrix). The probability for the photon to be
found on arm A (conditioned on successful post-selection of $\left\vert
\phi\right\rangle $) is given by $P(Y_{A}=1|\phi)=1$, and therefore on arm B \ $P(Y_{B}=1|\phi)=0$\ ( $Y_{i}=0,1$ denotes the
eigenvalues of $\widehat{Y}_{i}$): a non-destructive intermediate measurement
of the position degree of freedom will always find the photon on arm A (5).
However, if the photon polarization in the diagonal basis $\{\left\vert
\nearrow\right\rangle ,\left\vert \searrow\right\rangle \}$ on a particular
path $i$ i.e., $\widehat{X}_{i}$ \ is measured, instead of the position,
there is a non-zero probability to find a photon on arm B with diagonal
$\left\vert \nearrow\right\rangle $ or anti-diagonal $\left\vert
\searrow\right\rangle $ polarization. One can further specifically couple
$\widehat{X}_{i}$ to a qubit pointer whose excited state detects a quantity we
may call the \textquotedblleft net diagonal polarization\textquotedblright%
\ (defined for an arbitrary state $\alpha\left\vert \nearrow\right\rangle
+\beta\left\vert \searrow\right\rangle $ \ by $\left\vert \alpha
-\beta\right\vert ^{2}$\ (5); such a pointer could be found
excited on arm B, never on arm A. Hence with our choices of pre and
post-selected states, a pointer \textquotedblleft triggers\textquotedblright%
\ (for example, its position shifts) only on arm A if the spatial degree of
freedom is measured, but on arm B if a different property like the diagonal
component of polarization degree of freedom is measured. There is nevertheless
no paradox: Bohr (6) and Wheeler (7) proscribed long ago the use of
counterfactual reasoning while attempting to account for the behavior of
quantum systems measured under different experimental conditions within a
single picture. The reason is that measurements disturb the system. Indeed if
only~ $\widehat{Y}_{A}$ is measured, the photon will always be found on arm A,
but if only~ $\widehat{X}_{B}$ is measured it is not possible to ascribe a
property to the spatial position of the same photon on arm A had~ $\widehat
{Y}_{A}$ been measured. When $\{\widehat{X}_{B},\widehat{Y}_{A}\}$ \ are
measured jointly, the system coherence is disturbed, and we will obtain with
equal probabilities either the photon position on arm A or the photon's
diagonal polarization on arm B.

In order to keep the coherence essentially intact while jointly detecting the 
spatially separated properties on each arm for a single photon in the same run of the 
experiment,
minimally disturbing intermediate interactions need to be implemented. This is
the objective of our experiment whose results are reported below. For this
purpose, we use what are known as weak measurements (8) wherein very weak
couplings are combined with pre and post-selected states (as defined in Figure
1). In this situation, the shift of a pointer weakly coupled to a system
observable~ $\widehat{S}$ is proportional to the real part of a quantity known
as the weak value $S^{w}=\left\langle \phi\right\vert \widehat{S}\left\vert
\psi\right\rangle /\left\langle \phi\right\vert \left.  \psi\right\rangle .$
\ In the present setup, with the notation introduced previously, it follows that $Y_{A}^{w}=1,X_{B}^{w}=1$. These weak values imply
the following key feature: the pointer that is weakly coupled to the spatial
degrees of freedom (DOF) on arm A and the pointer that is weakly coupled to
the diagonal polarization DOF on arm B, both shift when a single photon passes
through the interferometer. In addition, a logical consequence (see Appdx B) of having these
weak values equal to unity is that $Y_{B}^{w}=0$ \ and $X_{A}^{w}=0$, so that weakly coupling the spatial DOF on arm B or the
diagonal polarization DOF on arm A has no effect on the respective pointers.
Thus, the pointers' motions resulting from the weak couplings can be
interpreted as reflecting the superposition of these two different photon
properties along spatially separated arms. The spatial separation of the
position degree of freedom from another property has amusingly been coined (9)
the \textquotedblleft Quantum Cheshire Cat\textquotedblright\ (QCC).

\begin{figure}[tb]
	\centering \includegraphics[scale=0.55]{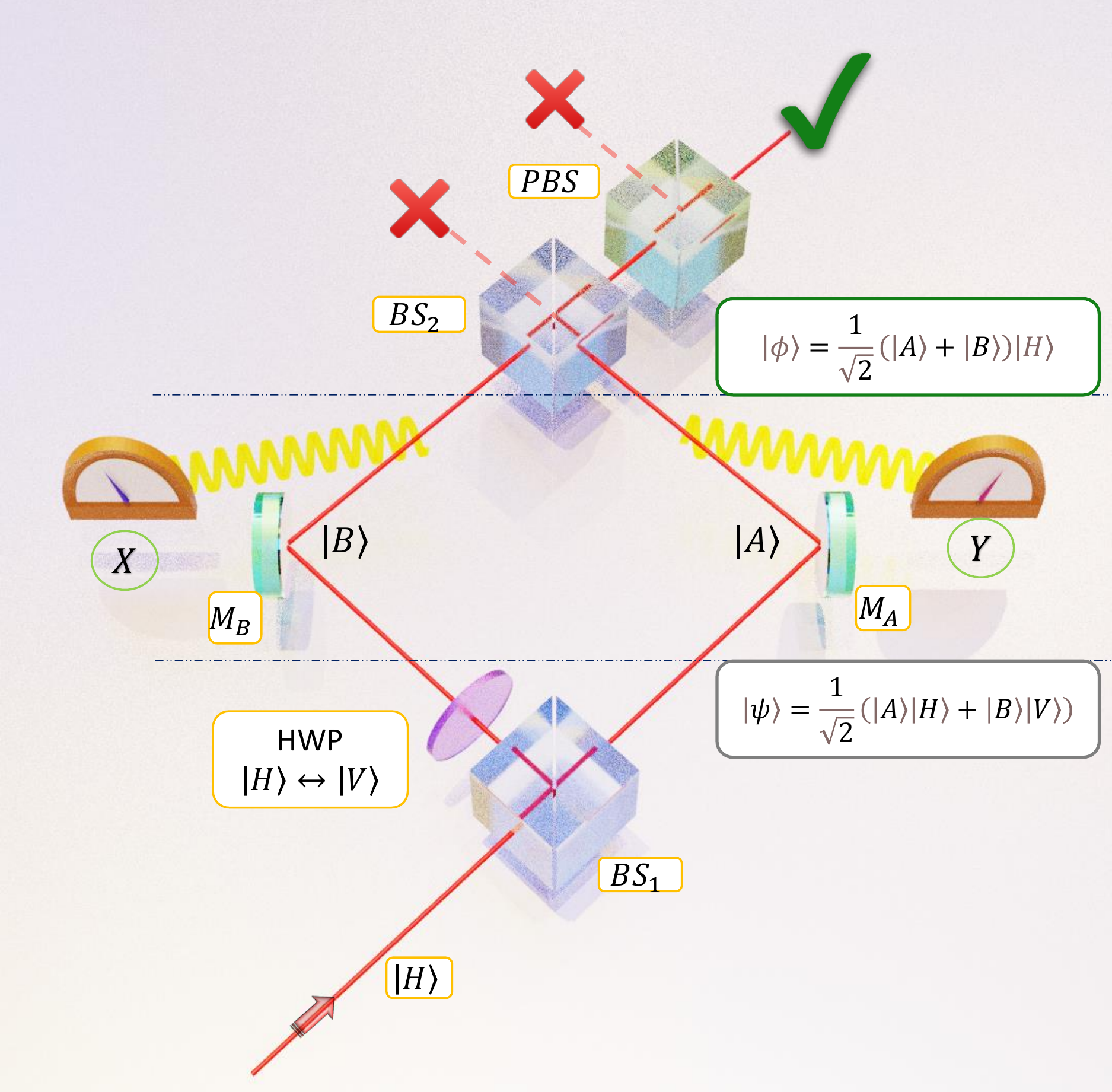}\caption{Pre and
		post selection in a Mach Zehnder Interferometer. The observables ~ $\widehat
		X$ and ~ $\widehat Y$ are coupled to the pointers at an intermediate time
		between pre-selection with state $\left\vert \psi\right\rangle $ \ and
		post-selection with state $\left\vert \phi\right\rangle $.}%
	\label{mtfig1}%
\end{figure}

Genuine weak measurements are generally delicate to implement experimentally
given that the coupling is weak and the experimentally measured quantities are
often of the same order of magnitude as the experimental errors for certain
choices of pre and post-selection. Several experiments in the last 15 years
have measured weak values and their ramifications (10-16). Very often
however, weak values are inferred by combining distinct projective (strong
coupling) measurements. This has been the case in particular for past
interferometric experiments dealing with the QCC, in which the weak values
were computed from the difference in the intensities obtained employing
set-ups with different experimental arrangements, as opposed to measurements
of pointer shifts arising from minimally disturbing interactions. In several
experiments (17-19) such intensity differences were obtained employing set ups
with or without absorbers present, and polarization or spin rotators inserted.
A more recent work reconstructed weak values from pointer shifts by
interpolating from results obtained using different interaction strengths
(20). However, such methods preclude the joint observation of different
properties in the same run (21). In order to observe the superposition of
spatially separated properties, it is crucial to implement non-destructive
minimally perturbing measurements on the same quantum particle along both arms.

In this work, employing the experimental architecture shown in Figure 2
with a single photon source, we perform joint weak measurements of
$\widehat{Y}_{A}$, the spatial DOF of the photon in arm A, and of $\widehat
{X}_{B}$, the diagonal component of polarisation DOF in arm B, in the same run
of the experiment i.e. without any change in experimental settings between pre
and post-selection. The photon is prepared in the pre-selected state
$\left\vert \psi\right\rangle $\ and weakly interacts with the optical
elements associated with the observables to be measured inside the
interferometer. The centre of the transverse spatial profile of the photon is
used as the pointer. \ For the interaction involving the spatial DOF, we use a
tilted glass plate that causes a vertical shift of the beam. The coupling of
$\widehat{X}_{B}$ \ is made with two beam displacers that cause a horizontal
shift in the pointer after the post-selection. A polarizing beam-splitter at
the output port post-selects the photon to the state $\left\vert
\phi\right\rangle $. Finally, the pointers' vertical and horizontal shifts are
measured. Each step of the experiment is detailed in the Appdx A.

\bigskip

\begin{figure}[tb]
	\centering \includegraphics[scale=0.45]{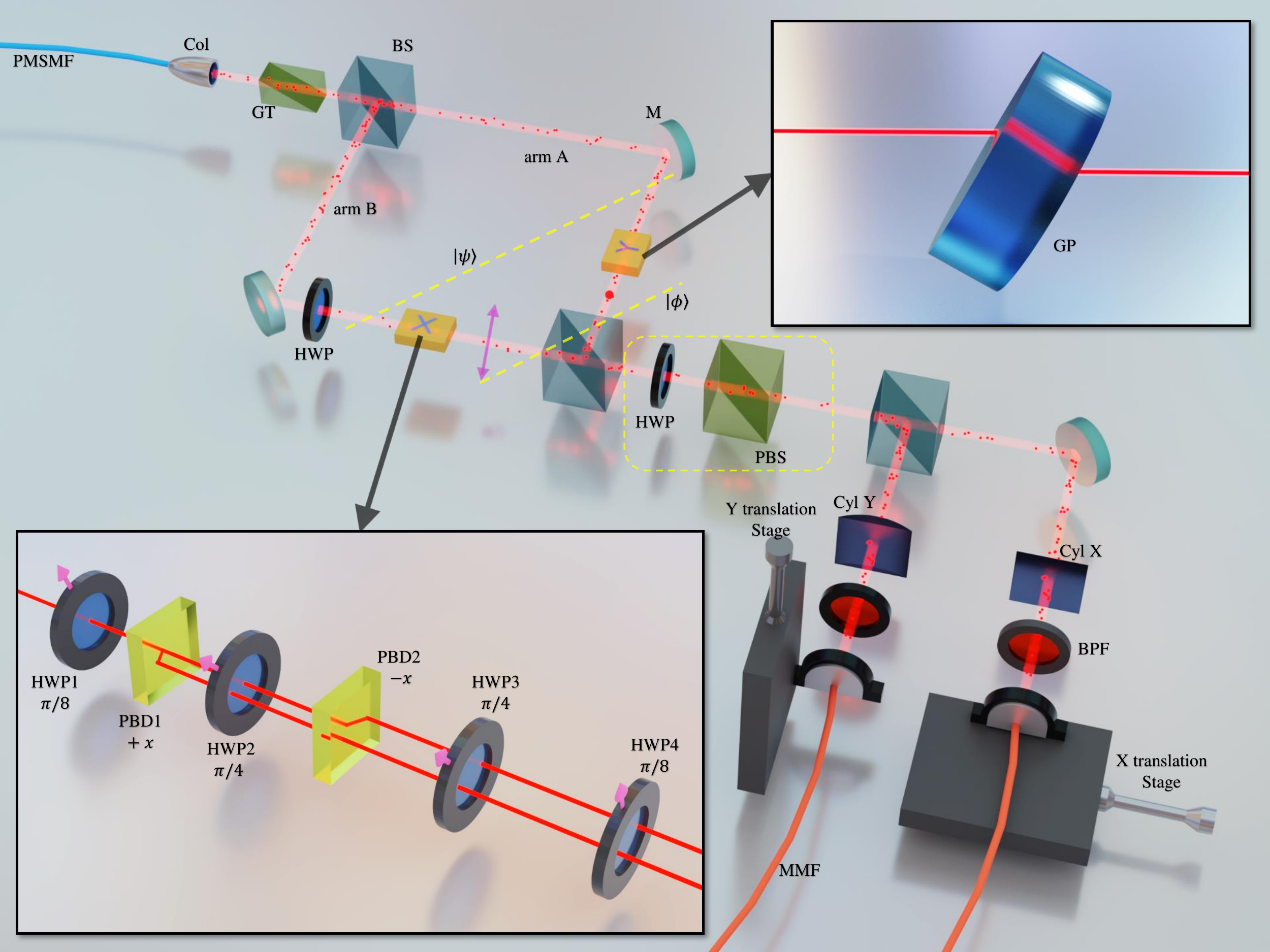}\caption{Experimental
		schematic shows the MZI with pre and post-selection. The angle of the
		post-selection half wave plate (HWP) is kept at zero degree. In arm B, the
		apparatus to measure the $\sigma_{1}$ \ polarization component is inserted,
		which displaces the beam along the horizontal (X). The glass plate (GP) in arm
		A makes the beam shift vertically (along Y). A 50:50 beam-splitter placed
		after the polarising beam splitter (PBS) causes the photons to randomly make
		their way to two multi-mode fibres (MMF). One of the fibres is moved along X
		to generate the horizontal profile, after the beam passes through a
		cylindrical lens that compresses the vertical transverse profile. The other
		MMF is moved along Y to generate the vertical profile, after the horizontal
		transverse profile of the beam is compressed using another cylindrical lens.
		These two interactions simultaneously occur and no setting is changed between
		pre and post selection during the data acquisition, thus enabling joint weak
		measurement of the two observables $\widehat{X}$ and $\widehat{Y}$.}%
	\label{mtfig2}%
\end{figure}

Being an interferometric experiment, steps have been taken to ensure maximum
coherence as well as proper phase stabilization. One of the key features that
we need to ensure in order to enable an unambiguous joint measurement is
perfect overlap of the beams from the individual arms in the absence of the
desired weak interaction. This involves a critical alignment procedure
involving measuring the undeviated beam positions, while including all
necessary interaction components. We also need to ascertain the values of the
exact pointer shifts in microns for both weak interactions that would
correspond to a weak value of 1, requiring further calibration (5). Ensuring
coherence and maximum visibility requires prior alignment with a pulsed laser
and a beam profiler camera before moving on to the heralded photon source. The
signal photon passes through the interferometric set-up while the photon in
the heralding arm is used to enable the measurement of coincidences at the
desired pre and post selection conditions. A crucial requirement in our
experiment is to ensure that we are performing joint measurements on the same
photon. The use of heralded photons in our experiment ensures the same. We
also show a measurement of cross correlation ($g_{2}$ measurement) in Appdx A to further substantiate the fulfillment of this requirement.

\begin{figure}[tb]
	\centering \includegraphics[scale=0.60]{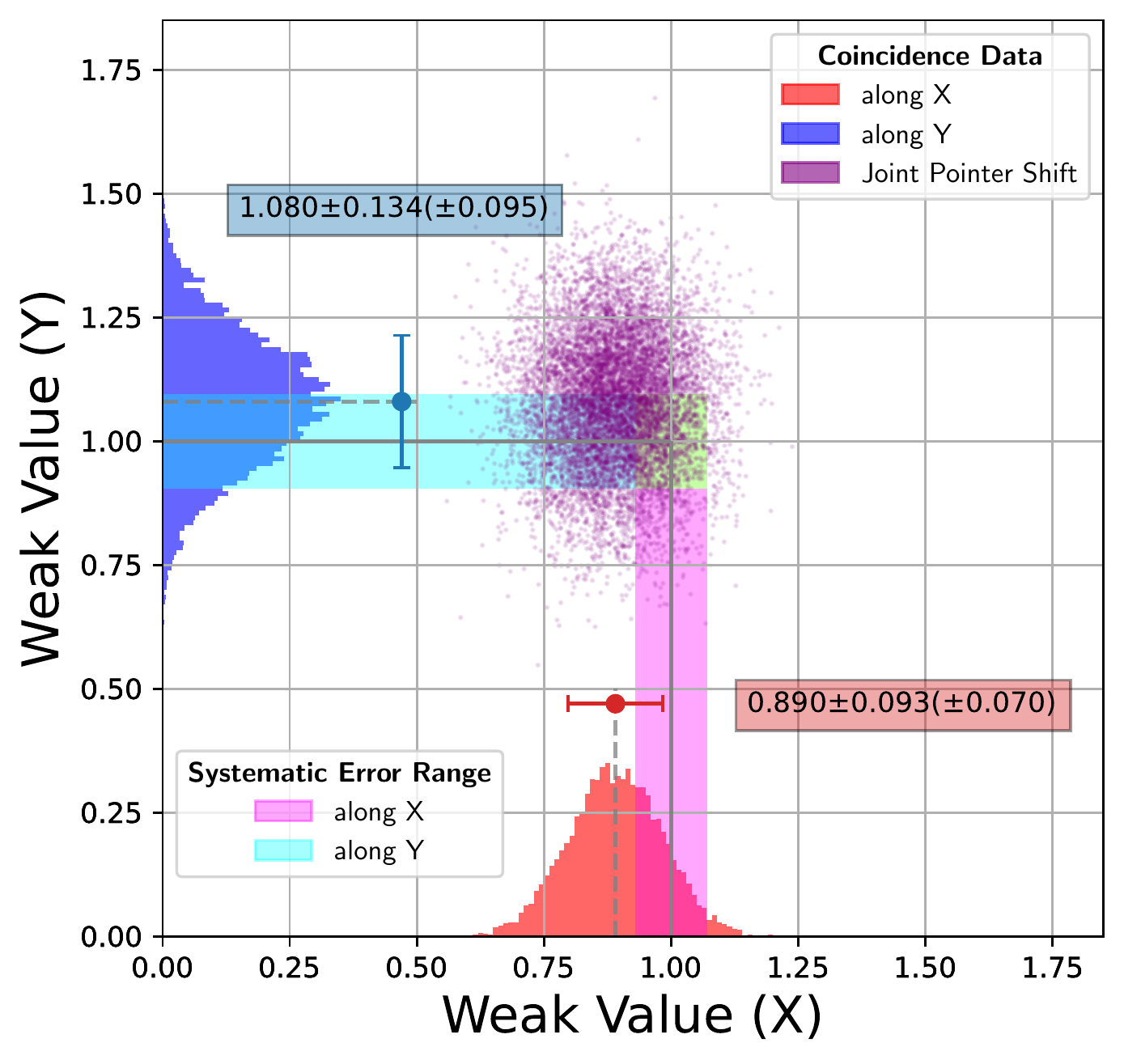}\caption{The
		coincidence distribution of the photons for $X^{w}$ \ is shown on the X axis
		through the red histogram plot. The mean and $1\sigma$ \ errors from the
		distribution are shown above the histogram while the values are mentioned
		alongside in the box. Similarly, the distribution for $Y^{w}$ is shown along
		the Y-axis through the blue histogram plot. Since both the pointer shifts
		occur concurrently and are measured jointly in the experiment (i.e., without
		changing anything between pre and post-selection), the overall shift of the
		pointer would be along the diagonal as inferred from the shifts on the
		projections along horizontal X and vertical Y. The distribution of these
		diagonal shifts, in terms of the weak values, are represented by the scatter
		plots. The pink and light blue bands represent the systematic error bands for
		the coincidence measurements (+/- 0.070 for $X^{w}$ \ and +/- 0.095 for
		$Y^{w}$ \ respectively). As can be seen, both measured weak values (0.890+/-
		0.093) for $X^{w}$ and (1.080+/- 0.134) for $Y^{w}$ \ lie well within the
		systematic error band of the experiment. Details of the error analysis are
		given in Appdx A.}%
	\label{mtfig3}%
\end{figure}

We show a representative result from our experiment in Figure 3. The figure
demonstrates the histogram of weak values obtained for $X^{w}$\ (component of
polarisation DOF) and $Y^{w}$\ (spatial DOF) measured jointly for the prepared
pre and post selected states given above. Experimentally we reconstruct the
transverse profile of the beam along the horizontal (X) and vertical (Y)
directions as mentioned in Figure 2 and measure the pointer shifts associated
with the two interactions  $\widehat{X}_{B}$ and  $\widehat{Y}_{A}$ from the
positions of the centres of the respective profiles. From $10^{4}$ \ such
profile reconstructions we obtain the pointer shift along X to be
$53.468\pm5.592$ $\mu m$ \ and the pointer shift along Y to be $56.809\pm
7.026$ $\mu m$, where the values represent the mean $\pm1\sigma$
\ error respectively. The weak values are then evaluated from the respective pointer
shifts; we obtain $X^{w}$  to be $0.89\pm0.09$  and $Y^{w}$ \ to be
$1.08\pm0.13$ \ respectively, along with attendant systematic error bands as
shown in Figure 3. \ The systematic error range estimates the drift in scale
with respect to which the weak values are computed. This is primarily caused
by beam pointing fluctuations as well as the drift in the centre of the beam
due to acoustic and thermal response of the optomechanical components. Details
of our data acquisition statistics as well as detailed error analysis (both
statistical as well as systematic errors) are discussed in Appdx A.

These results are, to our knowledge, the first direct joint measurement of the
weak values of different observables of a single quantum particle in distinct
spatial regions. We have experimentally shown that a quantum pointer on one
arm of the interferometer detects the spatial DOF of a photon in the chosen
pre and post-selected states; at the same time, a pointer in the other arm
detects the diagonal component of polarization DOF of the same photon. Our
approach could pave the way to develop technologies implementing distinct
interactions with different degrees of freedom of the same quantum system in
different spatial regions, with minimal mutual perturbations. It should be
worth exploring the application of the techniques used in our photonic
experiment for demonstrating similar effects in massive particles by relying
on the currently developing~coherent atom-chip Stern-Gerlach interferometry
(22) and thereby demonstrate the effect similar to the one shown in our
photonic experiment. This~could provide a potentially interesting dimension to
the studies aiming to test the applicability of fundamental quantum features
related to this work in the macroscopic regime. Our scheme could also be
applied to quantum information protocols, for instance, to share the state of
a qubit among spatially separated parties; a procedure for counterfactual
quantum communication based on the Quantum Cheshire Cat effect was proposed
very recently (23). Finally, we would like to conclude by raising the
following provocative question: by going beyond Bohr's dictum (24) that we have
no right to speak about what a photon does within an interferometer, can the
effect shown in our experiment be interpreted as refining Bohr's principle of
wave-particle complementarity? This is motivated by noting that in this
experiment, the observables corresponding to particle-like properties of a
single photon within each arm of the interferometer seemingly exhibit a
wave-like superposition inside the interferometer. Of course, such a question
needs to be formulated more precisely and revealing its full conceptual import
could be a stimulating line of future study.

\section*{Acknowledgments}
US acknowledges partial support provided by the Ministry of Electronics and
Information Technology (MeitY), Government of India under grant for Centre for
Excellence in Quantum Technologies with Ref. No. 4(7)/2020 -- ITEA and
QuEST-DST project Q-97 of the Govt. of India. DH and US would like to
acknowledge partial support from the DST-ITPAR grant
IMT/Italy/ITPAR-IV/QP/2018/G. DH also acknowledges support from the NASI
Senior Scientist fellowship. We thank R. Chatterjee, S. Chatterjee, K.
Joarder, Meena MS and Hafsa Syed for technical assistance.

\vspace{1.5cm}
\textbf{References}
\begin{enumerate}

	\item J. A. Wheeler, in Quantum Theory and Measurement, J. A. Wheeler, W. H.
	Zurek, Eds. (Princeton Univ. Press,  Princeton, NJ, 1984), pp. 182--200. 
	
	\item V. Jacques et al, Experimental realization of Wheeler's delayed-choice
	gedanken experiment.~\emph{Science}~315,  966 (2007). 
	
	\item Y. Aharonov, P. G. Bergmann, and J. L. Lebowitz, Time Symmetry in the
	Quantum Process of Measurement, Phys.  Rev.~134, B1410 (1964). 
	
	\item R. E. George et al, Classically undetectable wavefunction collapse, PNAS
	110 (10) 3777-3781 (2013). 
	
	\item See Appendix.

	\item N. Bohr, On the notions of causality and complementarity, Dialectica, 2:
	312-319 (1948). 
	
	\item J. A. Wheeler, How Come the Quantum?, Annals of the New York Academy of
	Sciences, 480: 304-31 (1986). 
	
	\item Y. Aharonov, D. Z. Albert, and L. Vaidman, How the result of a
	measurement of a component of the spin of a  spin-1/2 particle can turn out to
	be 100, Phys. Rev. Lett. 60, 1351 (1988). 
	
	\item Y. Aharonov~\emph{et al,}~Quantum Cheshire Cats,~\emph{New J.
		Phys.}~15~113015 (2013). 
	
	\item G. J. Pryde et al, Measurement of Quantum Weak Values of Photon
	Polarization, Phys. Rev. Lett. 94, 220405 (2005). 
	
	\item O. Hosten and P. Kwiat, Observation of the Spin Hall Effect of Light via
	Weak Measurements, Vol 319, Issue 5864,  pp. 787-790 (2008). 
	
	\item S. Kocsis et al, Observing the Average Trajectories of Single Photons in
	a Two-Slit Interferometer, Science, Vol  332, Issue 6034, pp. 1170-1173
	(2011). 
	
	\item J. S. Lundeen et al, Direct measurement of the quantum wavefunction,
	Nature 474, 188--191 (2011). 
	
	\item G. Nirala et al, Measuring average of non-Hermitian operator with weak
	value in a Mach-Zehnder interferometer,  Phys. Rev. A 99, 022111 (2019). 
	
	\item R. Ramos et al. Measurement of the time spent by a tunnelling atom
	within the barrier region. Nature 583, 529--532  (2020). 
	
	\item Y. Pan et al. Weak-to-strong transition of quantum measurement in a
	trapped-ion system. Nat. Phys. 16, 1206--1210  (2020). 
	
	\item T. Denkmayr et.al., Observation of a quantum Cheshire Cat in a
	matter-wave interferometer experiment, Nature  Communications, vol. 5, 4492
	(2014). 
	
	\item D. P. Atherton, G. Ranjit, A. A. Geraci and J. D. Weinstein, Observation
	of a classical Cheshire cat in an optical  interferometer, Optics Letters,
	vol. 40, no. 6, pp. 879-881, (2015). 
	
	\item J. M. Ashby, P. D. Schwarz and M. Schlosshauer, Observation of the
	quantum paradox of separation of a single  photon from one of its properties,
	Phys. Rev. A, vol. 94, p. 012102, (2016). 
	
	\item Y. Kim et. al, Observing the quantum Cheshire cat effect with
	noninvasive weak measurement, NPJ Quantum  Information, vol. 7, 13 (2021). 
	
	\item Q. Duprey et al, The Quantum Cheshire Cat effect: Theoretical basis and
	observational implications, Annals of  Physics, vol. 391, pp. 1-15, (2018).

	\item Y. Margalit et al, Realization of a complete Stern-Gerlach
	interferometer: Toward a test of quantum gravity, Sci.  Adv. 7, 22, eabg2879
	(2021). 
	
	\item Y. Aharonov, E. Cohen and S. Popescu, Nature Comm., vol. 12, p. 4770,
	(2021). 
	
	\item N. Bohr, in Albert Einstein: Philosopher-Scientist, ed. P. A. Schilpp
	(Library of Living Philosophers, 1951) p.  230.
\end{enumerate}

\pagebreak

\setcounter{figure}{0} \renewcommand{\thefigure}{A-\arabic{figure}}

\appendix

\section{Experimental Details}

\subsection{Experimental Architecture}

\textbf{Setting up the Mach-Zehnder interferometer}

The schematic of the experimental setup is given in Figure A-1. The stream of
single photons of wavelength 810 nm (bandwidth about 2 nm) is made incident on
the Mach Zehnder Interferometer (MZI) from a polarization maintaining single
mode fibre PMSMF$_{s}$\ [PM780-HP, Thorlabs] to minimize pointing fluctuations
of the beam. A suitable collimating lens COL [F240FC-780, Thorlabs] is used to
get a beam size of about 1.5 mm and minimize divergence. The beam is passed
through a Glan-Thompson polarizer GT [GTH5-B, Thorlabs] to ensure the
polarization is horizontal with a high degree of purity before entering the
50:50 beam splitter $BS_{1}$\ [BS014, Thorlabs]. In order to ensure that the
path difference between the two arms of the MZI is negligible compared to the
coherence length of the stream of detected photons, we use a corner cube
retroreflector CCR [PS976M-B, Thorlabs] mounted on an actuator [ZST225-B,
Thorlabs] attached to a 3D translation stage in one of the arms (here in arm
B). The actuator is adjusted so that the path difference is ensured to be
within $\sim2$ microns. The CCR is used instead of a mirror assembly to avoid
angular beam deviation upon translation. The CCR is attached to a piezo
(osi-stack) which is used to stabilize the phase difference between the two
arms of the MZI described later. However, since the CCR itself creates an
additional path difference, this is macroscopically compensated using the
three mirrors $M_{T_{1}},M_{T_{2}}$ and $M_{T_{3}}$\ [NIR 5102, Newport] in
the A arm of the MZI. \ The CCR, however, introduces ellipticity in the
polarization of the beam in arm B. This is corrected using a half-wave plate
$\mathit{HW}P_{c}$\ [WPA03-H-810, Newlight Photonics] followed by quarter-wave
plate $\mathit{QW}P_{c}$\ [WPA03-Q-810, Newlight Photonics] and the
polarization is made vertical. An additional polarizing beam splitter
$\mathit{PB}S_{T}$ \ \ is introduced after the three mirrors in arm A to
further purify the polarization. The second beam splitter $BS_{2}$ [BS014,
Thorlabs] \ is fixed at the intersection point of the two beams and oriented
to roughly ensure collinearity of the MZI. For fine alignment, the tip tilt
degrees of freedom of the mirror mounts \ are used to ensure collinearity and
the overlap of the beams \ is ensured by translation of the CCR. The overlap
of beams \ is ensured down to 5 microns (although the precision of measuring
the centre is much higher, at 0.1 microns, the spatial noise in the beam
limits the accuracy) and the angle between the \ two beams emerging out of the
second beam splitter ($BS_{2}$) \ is ensured to be less than $10^{-5}$
\ radians (the actual value may be around $10^{-7}$ \ radians as estimated
from the fringe stability of the interferometer). A half wave plate
$\mathit{HW}P_{\mathit{post}}$\ [WPZ0-200-L/2-810, Castech] mounted in a
motorized rotation mount [PRM1/MZ8, Thorlabs] is placed before a polarising
beam splitter  ($\mathit{PB}S_{\mathit{post}}$ at one of the exit ports of the
second beam splitter ($BS_{2}$). The post selected state changes with the
change in the angle of $\mathit{HW}P_{\mathit{post}}$. In addition to the
angle of interest, two more post selection angles are chosen to measure the
reference pointer positions associated with the eigen values 0 and 1
respectively. \ This enables evaluating the weak value for a particular post
selection from the measured pointer shift.

\textbf{Pre- and post-selection}

The pre-selected state is prepared after the compensating waveplate
$\mathit{QW}P_{c}$ on \ arm B and after \ $\mathit{PB}S_{T}$ \ in arm A as
depicted in Figure A-1. The post-selected state is obtained by back evolving
the transmitted component of the post-selection PBS ($\mathit{PB}%
S_{\mathit{post}}$) to a time before the second beam splitter $BS_{2}$ \ of
the MZI. Such a state, which is thus guaranteed to be transmitted in the
\ $\mathit{PB}S_{\mathit{post}}$\ , is given by $\left\vert \phi
(\theta)\right\rangle =\frac{1}{\sqrt{2}}\left(  \left\vert A\right\rangle
+\left\vert B\right\rangle \right)  {\otimes}S(\theta)\left\vert
H\right\rangle $. \ Here $S(\theta)$ \ is the Jones matrix for the HWP whose
fast-axis is orientated at an angle $\theta$ \ from the horizontal in the post
selection. \ 

\textbf{Ensuring Coherence}

A band pass filter $\mathit{BP10}$\ [FB810-10] followed by another
$\mathit{BP3}$\ [LL01-810-25] are used after $\mathit{PB}S_{\mathit{post}}$
\ to prevent light of other wavelengths from being detected and also to narrow
the linewidth down to at most 3 nm. The bandwidth of the filter determines the
minimum coherence length. \ Visibility of the interference upon a suitable
(diagonal, at $\theta=22.5^{{}^{\circ}}$) polarization post-selection is
measured as a function of the path length difference by moving the actuator on
which the CCR is mounted. Finally, the actuator is left at the position where
maximum visibility is obtained.

\begin{figure}[h]
\centering \includegraphics[scale=0.299]{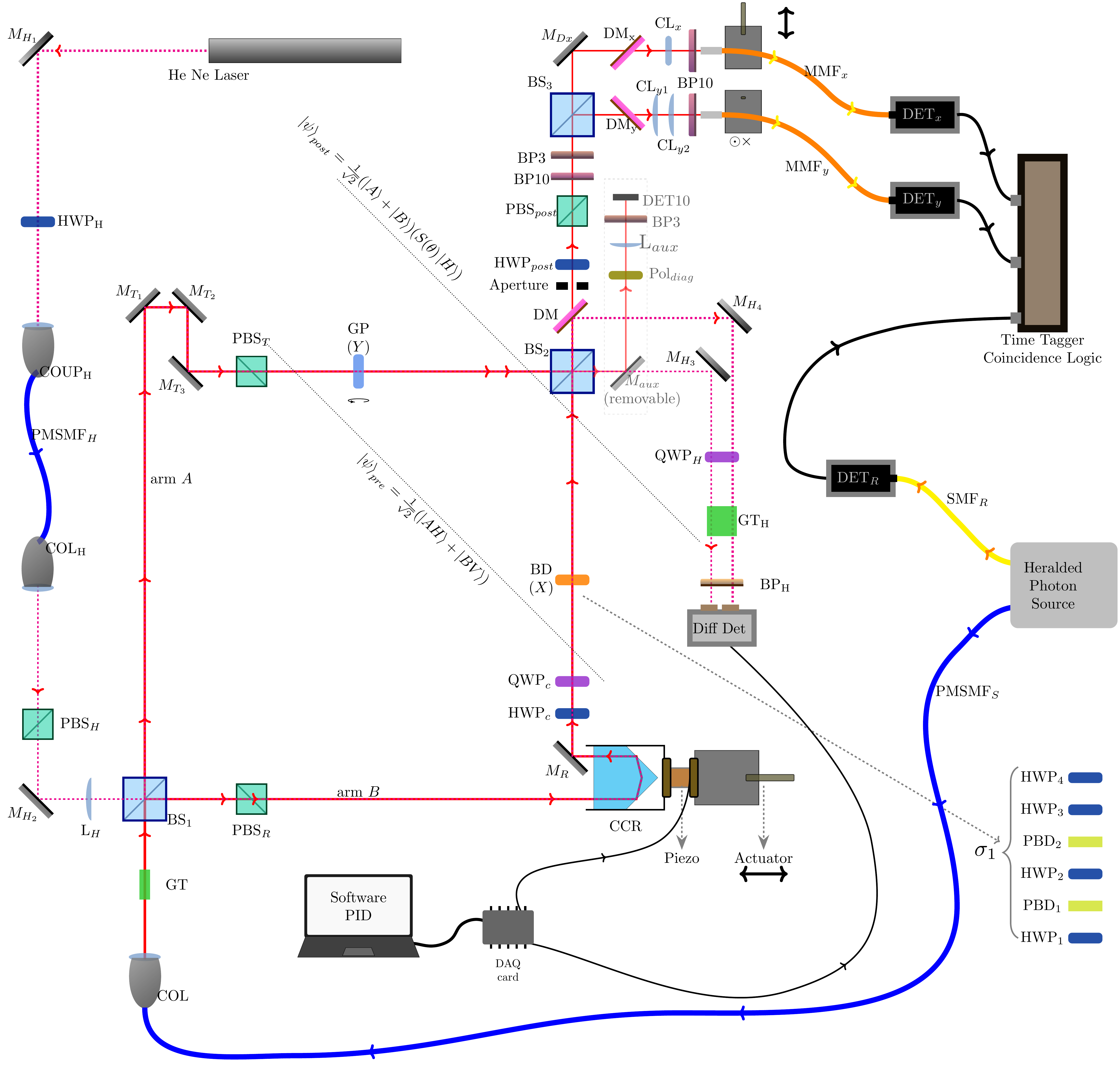}\caption{Experimental
setup}%
\label{mafig1}%
\end{figure}

\textbf{Phase Stabilization}

The relative phase between the two arms of the interferometer can drift due to
mechanical and acoustic vibrations and therefore the CCR needs to be moved
accordingly in order to maintain a constant phase relationship between the two
arms. This is achieved by using the piezo (attached to the CCR) which
contracts or expands depending on the voltage provided to it, thus causing the
CCR to move. \ For this, a PID algorithm is implemented on a computer along
with a DAQ card [USB-6003] which is used to generate and receive voltages.
\ The phase shift in the interferometer is monitored by measuring the
difference in power in the two output ports of the second beam splitter
$BS_{2}$ \ \ for a Helium Neon laser (He Ne) beam of wavelength 633 nm,
inserted into the MZI from the other input port of the first beam splitter
($BS_{1}$). This beam is mostly blocked by the Dichroic Mirror \ $\mathit{DM}$
[DMLP735, Thorlabs] placed after \ $BS_{2}$ \ in addition to the band pass
filters (BP10 and BP3). Before the differential detector (Diff Det), \ the
polarizations of the He Ne beams coming from two output ports of $BS_{2}$
\ are transformed into circular basis using quarter-wave plate $\mathit{QW}%
P_{H}$ \ and then projected to horizontal polarization using $GT_{H}$ \ to
achieve maximum visibility. The differential intensity signal obtained from
\ the 633 nm beam is calibrated with the intensity of the 810 nm laser
measured at the other output port (at which the post selection is not
performed) of \ $BS_{2}$ \ \ with diagonal polarization projection using
photodetector [DET10/M, Thorlabs] as a function of voltage given to the
\ piezoelectric transducer. When the path difference is within one wavelength,
these two signals become almost linear enabling the usage of the differential
intensity signal \ to stabilize the phase difference within low uncertainty.

\textbf{Weakly coupling the spatial DOF using a tilted glass plate}

A parallel window GP [WG41010-B, Thorlabs] is placed in arm A and is tilted to
cause a vertical shift of $\sim50$ microns in the beam. Tuning the angle of
tilt of the glass plate, the shift in the beam can be controlled. If the shift
were more than the beam width of 1.9 mm, observing this shift after the
post-selection would have indicated whether the photon came from arm A or arm
B. The shift of $\sim50$ microns being much less than the beam width ensures
that the observable $(\widehat{Y}_{A})$ \ is weakly coupled.

\textbf{Weakly coupling the diagonal component of polarisation DOF using a
composite beam displacer}

A polarising beam displacer (PBD) allows the ordinary component of
polarization of the beam incident on it to pass through without any deviation
and causes a lateral shift (depending upon its thickness) in the path of the
extraordinary component of polarization of the incident beam. Thus, the
operation of a PBD can be considered as $\sigma_{3}$ \ measurement operation.
The $\sigma_{1}$ \ measurement operator can be constructed from the
$\sigma_{3}$ \ measurement operation. Here the \ $\sigma_{3}$ \ measurement
operator is created using two beam displacers $\mathit{PB}D_{1}$ \ and
$\mathit{PB}D_{2}$\ [PDC 12005, \ Newlight Photonics] oriented in such a way
that one of them \ causes $\sim50$ microns shift in the extraordinary
component \ along one direction (say +X) and the other shifts the
extraordinary component by $\sim50$ microns in the opposite direction (-X)
horizontally. A half wave plate $\mathit{HW}P_{2}$\ (fast axis at $\pi/4$) is
inserted between the two beam displacers so that the extra-ordinary beam for
the first \ PBD becomes the ordinary component for the second PBD and vice
versa. \ This is followed by another half wave plate $\mathit{HW}P_{3}$\ (with
fast axis at $\pi/4$) so that the ordinary and the extra-ordinary polarized
beams have the same phase in the description of the $\sigma_{3}$ \ operator.
\ The whole $\sigma_{3}$ \ measurement operator is placed between two half
wave plates $\mathit{HW}P_{1}$ \ and $\mathit{HW}P_{4}$ [WPA03-H-810] with
their fast axes oriented at angle $\pi/8$ \ to realise the $\sigma_{1}$ \ operator.

\textbf{Alignment procedure for joint observation}

Ideally, the zero reference of the beam, on a 2D plane after the
post-selection would be at the position where both the beams from arm A and
arm B merge. They are expected to be overlapping and collinear at the
detection plane. However, when components for the weak interaction with the
$\sigma_{1}$ \ polarization component are inserted in arm B, the beam
displacers ($\mathit{PB}D_{1}$ and $\mathit{PB}D_{2}$) need to be tilted in
order to adjust the phase shift between the emergent e-ray and o-ray so that
in the limit of the beam displacement between them going to zero, the
evolution operator due to the interaction remains identity. Due to the tilt as
well as a slight angular deviation of the beam from components, the beam in
arm B gets refracted after the components are inserted. This causes a slight
non-collinearity and translation of the beam. \ The change in collinearity is
very small and can be adjusted by maximizing the visibility when post
selection is either diagonal or anti-diagonal. This is achieved by the
tip/tilt of the mirror $M_{R}$. The displacement between the diagonal and
anti-diagonal components is fixed by the alignment in components
($\mathit{PB}D_{1}$ and $\mathit{PB}D_{2}$) for the $\sigma_{1}$
\ interaction. However, they should ideally be on either side, equidistant
from the centre of the beam emerging from arm A. One way to verify this would
be to measure the beam centre position from arm A with arm B blocked and vice
versa. However, when arm A is blocked, the pre-selected polarization in arm B
is orthogonal to the post-selected polarization. Due to the displacement
between diagonal and anti-diagonal components, we have the destructive
interference pattern after post-selection as shown in Figure A-2.

\begin{figure}[t]
\centering \includegraphics[scale=0.6]{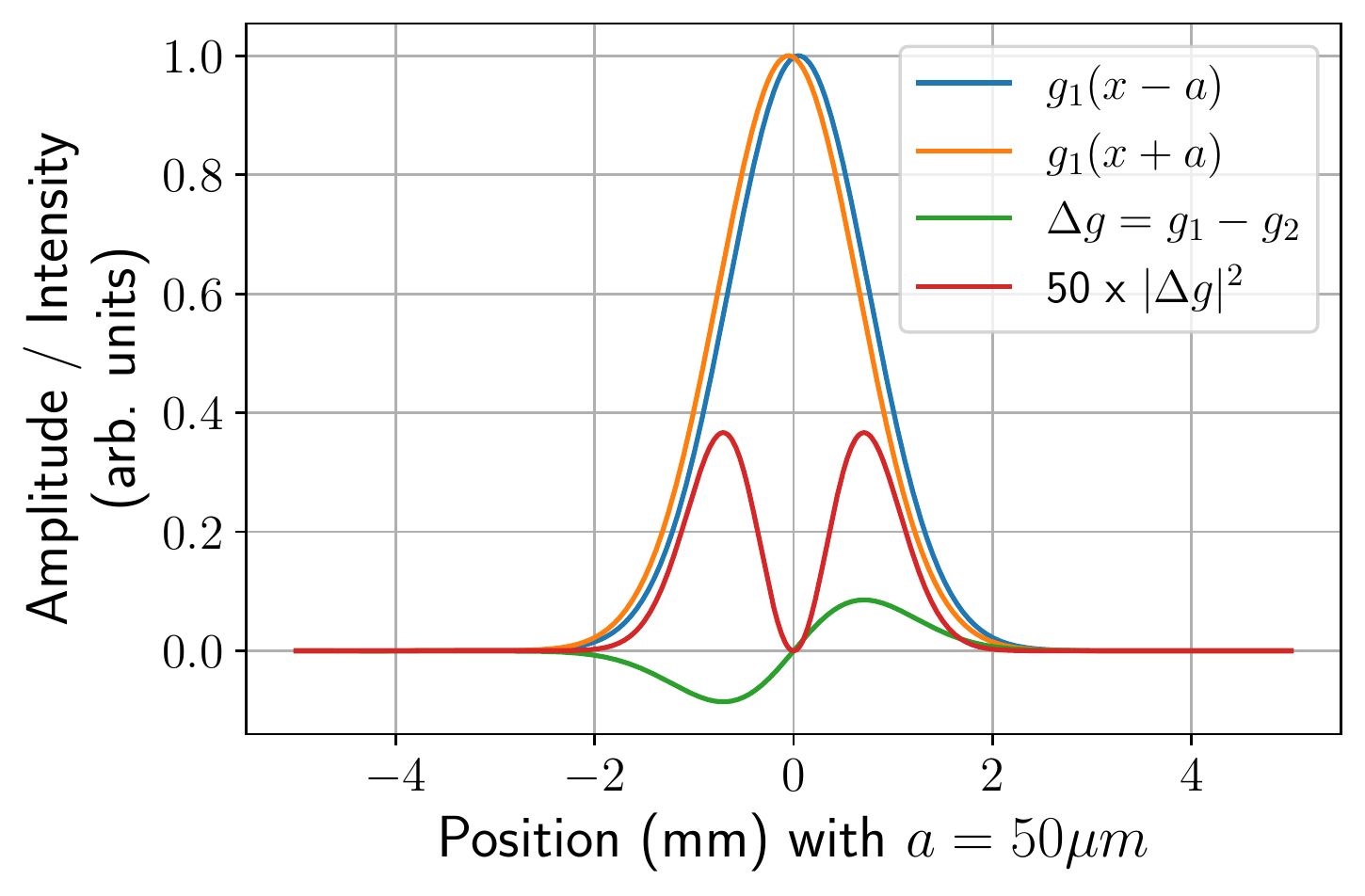}\caption{Destructive
Interference intensity pattern (shown after multiplication by 50 in red) is
obtained when two Gaussians $g_{1}$ \ and $g_{2}$ \ are displaced in opposite
direction by the displacement $a$. The centroid of this interference pattern
has more error due to the presence of noise. The centroid of the pattern is
also very sensitive to the phase difference between the two Gaussian beams.}%
\label{mafig2}%
\end{figure}

The centroid of the destructive interference pattern tends to be error prone.
Therefore, utilising the fact that the centres of the beams in arms A and arm
B coincide, one needs to use the resultant interference pattern formed when
the post-selection is diagonal and anti-diagonal. From theory, these two
displacements need to be symmetric about the centre of \ the beam of arm A.
This is achieved by translating the CCR and measuring the resultant centroid
at three post-selection configurations. Once the above alignment is ensured,
the glass plate in arm A which was back aligned, is now tilted so that it
causes the desired shift along the vertical direction when arm B is blocked.
All the above steps are performed with pulsed laser and data is acquired with
a beam profiler. The 2D centroid of the beam is scaled to the weak value. The
scaling can be achieved by subtracting the ensured zero position as mentioned
earlier and scaling it with the pointer shift corresponding to the eigenvalues
with individual arms blocked. This can be done with beam profiler data as
blocking and insertion of components are all performed with beam profiler as
the monitoring tool for the pulsed laser source. However, once the measuring
device is changed, the information about absolute position in different
configurations is lost. The beam profiler is then taken out of the setup and
the multimode fibres mounted on a translation stage are used. The weak value
is measured at the three post-selection angles to ensure that the alignment of
multi-mode fibres is consistent with the beam profiler data. Once the above is
ensured, the pulsed laser beam is replaced with stream of single photons
coming from a single photon source to measure singles and coincidences. Here,
we scale the pointer shift to weak values by subtracting the pointer shift at
post-selection HWP angle $45^{{\circ}}$\ \ (corresponding to the zero weak
value). Similarly, displacement by weak value 1 happens when post-selection
HWP angle is $90^{{\circ}}$.

\bigskip

\bigskip

\subsection{Results}

\textbf{Data Acquisition}

In order to subsequently use a single photon source, a 100 fs pulsed laser
[Mira 900 D Ti:Sapphire, rep rate: 76 MHz] is used with average power of about
2 mW to \ ensure good alignment. Although the coherence length of the pulsed
laser is lower (than CW), it is slightly enhanced by the use of the bandpass
filter so that the pulsed laser has almost the same central wavelength and the
bandwidth as the single photon source to be used later. A beam profiler
[Dataray UCD -15] is placed after the bandpass filter. The post-selection HWP
angle \ is rotated to the angle $67.5^{{\circ}}$ \ that creates the
destructive interference profile along the diagonal. The phase difference is
adjusted to make this destructive interference pattern as symmetric as
possible along the diagonal. The slight spatial noise in the beam profile
i.e., deviation from a perfect Gaussian profile typically would dominate the
destructive interference pattern. The overall power incident on the beam
profiler also reaches the minimum which is another indicator for the correct
phase difference to which the Mach Zehnder interferometer is locked. The post
selection HWP angle is rotated and 5 images are \ captured for each HWP angle.
The first order centroid of the 2D images is computed. The pointer shifts are
scaled to the weak value using the pointer shifts at $45^{{\circ}}$ \ and
$90^{{\circ}}$ \ as reference.

For the single photon source, the timing information along with the spatial
profile needs to be measured. The timing information ensures that the
contribution of multi-photon events to the pointer shift is negligible. For
this SPAD detectors [Tau SPAD-20, Pico quant] are used. The photons are
collected using a bare multimode fibre [M42L02, Thorlabs] with a core diameter
of 50 microns. These fibres ($\mathit{MM}F_{x}$ \ and $M\mathit{MF}_{y}$)
\ are moved in steps of 50 microns to sample the Gaussian profile. A width of
about 3 mm is covered with 61 points. \ Although the use of such fibres
averages the intensity over the 50 microns, the precision with the centroid is
much better than this and depends on the total number of photons collected. To
enhance the collection, a (combination of) cylindrical lens ($CL_{x}$,
$CL_{y_{1}}$ and $CL_{y_{2}}$) is used to compress one spatial dimension. The
multi-mode fibre tip is placed at the focus of the cylindrical lens and is
translated to obtain the spatial profile.

Since the objective is to jointly observe the two pointers, a 50:50 beam
splitter $BS_{3}$\ [BS014] is used to divide the beam. One beam is used to
reconstruct the horizontal profile (X) and the other beam is used to
reconstruct the vertical profile (Y). \ Since the beam splitter sends the
photon towards the horizontal or the vertical detector with inherent
randomness and there is no change of experimental settings within the
interferometer (i.e., between pre and post-selection), the horizontal and
vertical centres of the beam are jointly obtained for the ensemble of photons
for a given post-selected state. \ \ 

\textbf{Statistics and Error Analysis}

When the post selection HWP angle is set to $0^{{\circ}}$, 16 coincidence
readings per position of the MMF are taken. Any one of these 16 readings at a
particular position is chosen and it is repeated for all the 61 positions to
construct a Gaussian profile. Thus $16^{61}$ \ such possible profiles can be
constructed from the collected data. Out of that, $10^{4}$ \ such Gaussian
profiles are sampled and each such profile is fitted with a Gaussian function
to determine the centre. This gives the distribution of the position of the
pointer (say $X$) for the post-selected state $\left\vert \phi\right\rangle
$\ (when the $\mathit{HW}P_{\mathit{post}}$ \ angle is at $0^{{}^{\circ}}$).

Let us call the distribution of centres when the post-selection angle is
$45^{{\circ}}$ \ as $X_{0}$. The $10^{4}$ \ profiles for this \ post-selected
\ state \ $\left| \phi\left( 45^{{}^{\circ}}\right) \right\rangle $ \ are
\ sampled from datasets with 3 readings for each MMF position \ (amounting to
$3^{61}$ \ possible profiles).

Similarly, we have 3 readings per MMF position at the post-selection angle
$90^{{\circ}}$. The corresponding distribution of centres of the beam for the
post selected state $\left\vert \phi\left(  90^{{}^{\circ}}\right)
\right\rangle $ \ is, say $X_{1}$, which again is obtained from Gaussian fits
of $10^{4}$ \ profiles.

Since the profiles are drawn at random, we can create the distribution of weak
values by \ computing,%

\[
X^{w}=\frac{X-X_{0}}{\left\langle X_{1}-X_{0}\right\rangle }%
\]
The statistical random errors are represented by standard deviation of $X^{w}%
$. The random errors come from the numerator since we have taken the
expectation value in the denominator. This is to avoid spurious weak values if
some of the values in the denominator turn out to be zero by chance.

On measuring few profiles at a particular post-selection angle and then
changing the post-selection angle to observe a certain shift, it could so
happen that the observed shift is more or less than expected due to drift in
the position of the beam over time. This drift typically arises from the
effect of temperature, pressure and humidity on opto-mechanics and fibres
(mainly $\mathit{PMSM}F_{s}$). The beam drift is estimated in a separate
experiment where the transverse profile of the beam in arm A (with arm B
blocked) is repeatedly measured with the moving MMF for a long period. The
drift in centroid would represent the drift of the beam's centre over time.
The standard deviation of the centroid over time indicates the range of the
systematic error which changes the \ unit scale of the weak value.

\textbf{Single Photon Source Characteristics}

The experiment uses a heralded single photon source generated using
spontaneous parametric down-conversion of 405 nm pump beam in a type-II PPKTP
crystal in a collinear Sagnac geometry. As shown in Figure A-1 one stream of
photons (signal photons) is made incident on the MZI using a polarization
maintaining single-mode fibre ($\mathit{PMSM}F_{S}$). Although the source is
bright and singles count rate is in the MHz domain, photons with rates in the
kHz domain are detected with the multi-mode fibres ($\mathit{MM}%
F_{x},\mathit{MM}F_{y}$ shown in Figure A-1 due to the collection area of the
fibres. Each stream of photons (singles) is known to have Poissonian
statistics and only the heralded photons (coincidences) follow sub-Poissonian
statistics. \ The heralded cross-correlation function is defined for the SPDC
process as
\[
g_{2}\left(  \Delta\tau_{1},\Delta\tau_{2}\right)  =\frac{N(R)C\left(
S_{1}(\Delta \tau _{1}),S_{2}(\Delta \tau _{2})|R\right)  }{C\left(
S_{1}(\Delta \tau _{1})|R\right)  C\left(  S_{2}(\Delta \tau
_{2})|R\right)  }.
\]
Here, $N\left(  R\right)  $ \ is the number of photons in the reference (R)
detector \ $\mathit{DE}T_{R}$ \ in a given time duration (see Figure A-3).
\ $C\left(  S_{1}(\Delta \tau _{1}),S_{2}(\Delta \tau _{2})|R\right)
$ refers to the triple coincidences when $S_{1}$ \ is delayed by $\Delta
\tau_{1}(=\Delta\tau_{RS_{1}})$ \ and \ $S_{2}$ \ is delayed by $\Delta
\tau_{2}(=\Delta\tau_{RS_{2}})$ \ with respect to $R$. The coincidences
between $R$ \ with $S_{1}$ \ delayed by $\Delta\tau_{1}$ \ and between $R$
\ with $S_{2}$ \ delayed by $\Delta\tau_{2}$ \ is denoted by $C\left(
S_{1}(\Delta \tau _{1})|R\right)  $\ and $C\left(  S_{2}(\Delta \tau
_{2})|R\right)  $ respectively.

The $g_{2}(0,0)$ \ is about 0.038 thus implying that the stream of photons
from $S_{1}$ \ or $S_{2}$ \ when heralded with $R$ \ shows single-photon characteristics.

\begin{figure}[tb]
\centering \includegraphics[scale=0.39]{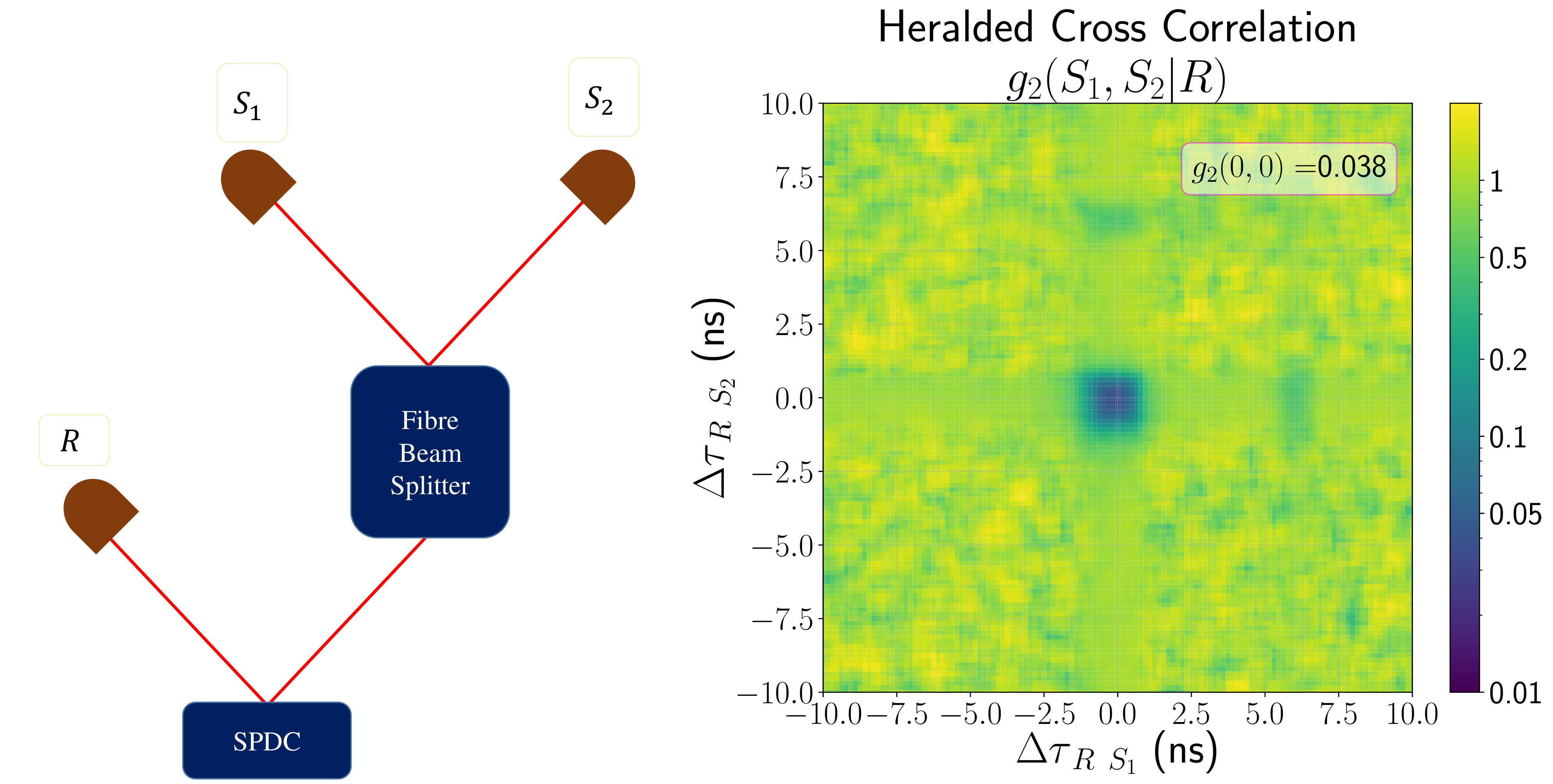}\caption{Heralded cross
correlation (pump power 20 mW, coincidence time window 312.5 ps).}%
\label{mafig3}%
\end{figure}

\section{Theoretical details}

\subsection{ Standard intermediate measurements in the Mach-Zehnder
interferometer}

\subsubsection{Setting}

We consider the Mach-Zehnder interferometer given in the main text and derive
the probabilities for strong intermediate measurements along the arms $A$ and
$B$. The pre and post-selected states were given repectively by
\begin{equation}
\left\vert \psi\right\rangle =\frac{1}{\sqrt{2}}\left(  \left\vert
A\right\rangle \left\vert H\right\rangle +\left\vert B\right\rangle \left\vert
V\right\rangle \right)  \label{pre}%
\end{equation}
and%
\begin{equation}
\left\vert \phi\right\rangle =\frac{1}{\sqrt{2}}\left(  \left\vert
A\right\rangle +\left\vert B\right\rangle \right)  \left\vert H\right\rangle .
\label{post}%
\end{equation}
$\left\langle r\right\vert \left.  A\right\rangle $ and $\left\langle
r\right\vert \left.  B\right\rangle $ are the spatial wavefunctions of the
photon along arms A and B ($r$ is the position), while $\left\vert
H\right\rangle $ and $\left\vert V\right\rangle $ correspond to horizontal and
vertical polarization.

The properties we are interested in are:

\begin{itemize}

\item the spatial projectors on arms A and B, represented by the observables
$\hat{Y}_{A}=\left\vert A\right\rangle \left\langle A\right\vert
\otimes\mathbb{\hat{I}} $ and $\hat{Y}_{B}=\left\vert B\right\rangle
\left\langle B\right\vert \otimes\mathbb{\hat{I}} $, having eigenvalues
$Y_{i}=0,1$ ($i=A,B$); note that
\begin{equation}
\hat{Y}_{A}+\hat{Y}_{B}=\mathbb{I}, \label{proj-c}%
\end{equation}
so that a positive measurement of $\hat{Y}_{A}$ (with eigenvalue $Y_{A}=1$)
is  also a negative measurement of $\hat{Y}_{B}$ (with eigenvalue $0$).

\item the diagonal polarization of the photon on each arm, represented by the
observables $\hat{X}_{A}=\left\vert A\right\rangle \left\langle A\right\vert
\otimes\sigma_{1}$ and $\hat{X}_{B}=\left\vert B\right\rangle \left\langle
B\right\vert \otimes\sigma_{1}$ where $\sigma_{1}$ is one of the Pauli
matrices with eigenstates
\begin{align}
\left\vert \nearrow\right\rangle  &  =\frac{1}{\sqrt{2}}\left(  \left\vert
H\right\rangle +\left\vert V\right\rangle \right) \\
\left\vert \searrow\right\rangle  &  =\frac{1}{\sqrt{2}}\left(  \left\vert
H\right\rangle -\left\vert V\right\rangle \right)  .
\end{align}
$\left\vert \nearrow\right\rangle $ and $\left\vert \searrow\right\rangle $
correspond to diagonal and anti-diagonal polarization. The eigenvalues of
$\hat{X}_{i}$ are $X_{i}=-1,1$ and $0$ (corresponding respectively to
anti-diagonal, diagonal and no polarization on arm $i$).
\end{itemize}

\subsubsection{Spatial degree of freedom}

Let us first consider a non-destructive measurement of the spatial
wavefunction when the photon is prepared in state $\left\vert \psi
\right\rangle $. If we measure the spatial degree of freedom on arm A, the
state after the intermediate measurement is projected to $\hat{Y}%
_{A}\left\vert \psi\right\rangle =\left\vert A\right\rangle \left\vert
H\right\rangle $ if the measurement is successful ($Y_{A}=1$). This happens
with a probability%
\begin{equation}
P(Y_{A}=1)=\mathrm{Tr}_{Y}\hat{Y}_{A}\mathrm{Tr}_{X}\left\vert \psi
\right\rangle \left\langle \psi\right\vert =\frac{1}{2}.
\end{equation}
The probability of post-selection after this successful measurement is then%
\begin{equation}
P(\left\vert \phi\right\rangle |Y_{A}=1)=\left\vert \left\langle
\phi\right\vert \left.  A\right\rangle \left\vert H\right\rangle \right\vert
^{2}=\frac{1}{2}.
\end{equation}
Since $P( \phi)=\left\vert \left\langle \phi\right\vert \left.  \psi
\right\rangle \right\vert ^{2}=1/4$, following Bayes' theorem the conditional
probability of finding the particle's spatial degree of freedom on arm A when
post-selection is succesful is \footnote{These  conditional probabilities are
known as the ABL\ rule, see Y. Aharonov, P. G.  Bergmann and J. L. Lebowitz,
Phys. Rev. B 134, 1410 (1964).}%
\begin{equation}
P(Y_{A}=1|\phi)=1.
\end{equation}
Given Eq. (\ref{proj-c}), the conditional probability of finding the position
degree of freedom on arm B given post-selection is therefore%
\begin{equation}
P(Y_{B}=1|\phi)=0.
\end{equation}
This can also be seen from the direct computation of $P(\left\vert
\phi\right\rangle |Y_{B}=1)=\left\vert \left\langle \phi\right\vert \left.
B\right\rangle \left\vert V\right\rangle \right\vert ^{2}=0$.

Let us introduce a pointer on arm A in state $\left\vert \xi_{x_{0}%
}\right\rangle _{A}$; its wavefunction is $\langle x|\xi_{x_{0}}\rangle
_{A}=\xi_{x_{0}}(x)$, where $x_{0}$ is the mean initial position of the
pointer. Assume the usual von-Neumann type coupling
\begin{equation}
H_{int}=\gamma(t)\hat{Y}_{A}\hat{P} \label{hint}%
\end{equation}
between the pointer's momentum $\hat{P}$ and the particle observable $\hat
{Y}_{A}$. $\gamma(t)$ is the coupling strength and let $g=\int\gamma(t)dt$
label the overall coupling strength taken over the interaction time. $g$ is
chosen such that the shifted pointer state $\left\vert \xi_{x_{0}
+g}\right\rangle _{A}$ is orthogonal to the initial state, i.e. $_{A}%
\left\langle \xi_{x_{0}+g}|\xi_{x_{0}}\right\rangle _{A}=0.$ Unitary evolution
brings the initial state%
\begin{equation}
\left\vert \Psi(t_{i})\right\rangle =\left\vert \psi\right\rangle \left\vert
\xi_{x_{0}}\right\rangle _{A} \label{init}%
\end{equation}
to the time evolved state after the interaction%
\begin{align}
\left\vert \Psi(t)\right\rangle  &  =e^{-ig\hat{Y}_{A}\hat{P}/\hbar}\left\vert
\psi\right\rangle \left\vert \xi_{x_{0}}\right\rangle _{A}\label{ecu0}\\
&  =\frac{1}{\sqrt{2}}\left(  \left\vert A\right\rangle \left\vert
H\right\rangle \left\vert \xi_{x_{0}+g}\right\rangle _{A}+\left\vert
B\right\rangle \left\vert V\right\rangle \left\vert \xi_{x_{0}}\right\rangle
_{A}\right)  , \label{ecu}%
\end{align}
where we have used the translation property of the operator $e^{-ig\hat{Y}
_{A}\hat{P}/\hbar}$. Postselecting the system to state $\left\vert
\phi\right\rangle $ [Eq.(\ref{post})] leaves the pointer in the shifted state
$\left\vert \xi_{x_{0}+g}\right\rangle $. This indicates that the postselected
photon has traveled through arm A. Similarly, we can place a pointer on arm B,
in the initial state $\left\vert \xi_{x_{0}}\right\rangle _{B}$. Eq.
(\ref{ecu}) is then replaced by%
\begin{equation}
\frac{1}{\sqrt{2}}\left(  \left\vert A\right\rangle \left\vert H\right\rangle
\left\vert \xi_{x_{0}}\right\rangle _{B}+\left\vert B\right\rangle \left\vert
V\right\rangle \left\vert \xi_{x_{0}+g}\right\rangle _{B}\right)
\end{equation}
After post-selection to state $\left\vert \phi\right\rangle $, the pointer
will be found in its initial state $\left\vert \xi_{x_{0}}\right\rangle _{B}$
(since the term $\left\vert B\right\rangle \left\vert V\right\rangle
\left\vert \xi_{x_{0}+g}\right\rangle _{B}$ is orthogonal to the post-selected
state): there is no displacement of the pointer, indicating that the
postselected photon has not traveled through arm B.\newline

The conclusion is that a standard non-destructive intermediate measurement of
the spatial degree of freedom with pre and post-selection states given by Eqs.
(\ref{pre}) and (\ref{post}) will always result in a pointer shift on arm A,
while the pointer on arm B remains in the initial (unshifted) state.

\subsubsection{Diagonal polarization degree of freedom}

While an intermediate measurement of the spatial degree of freedom (DOF) is
only detected on arm A, it is easy to see that measuring the diagonal
polarization DOF on arm B can result in detecting $\left\vert \nearrow
\right\rangle $ or $\left\vert \searrow\right\rangle $ with a probability
given by
\begin{equation}
P(X_{B}=\pm1|\phi)=\frac{1}{4}.
\end{equation}
So although the spatial DOF is never detected on arm B, a different property,
namely the diagonal component of the polarization DOF will be detected there
half of the runs. A pointer's momentum can be coupled to the diagonal
polarization observable $\sigma_{1}$ on arm A or B, similarly to the spatial
projection operator case (\ref{ecu0}) (with now two shifted states $\left\vert
\xi_{x_{0}\pm g}\right\rangle $ corresponding to $X_{i}=\pm1$).\newline

It is also possible to couple a qubit pointer to the diagonal polarisation of
the photon on a given arm, so that the interaction Hamiltonian reads
$H_{int}=\gamma(t)\sigma_{1}\sigma_{2}^{q}$ where $\sigma_{2}^{q}$ is a Pauli
matrix relevant to the qubit pointer and $g=\int\gamma(t)dt$ can be chosen so
that the initial qubit state $\left\vert 0_{q}\right\rangle $\ is rotated
conveniently. Such a qubit pointer will only be excited when placed in arm B,
as we will see below. Let us first examine the qubit behaviour after it gets
coupled to the system through $H_{int}$. Consider a polarized state in the
diagonal basis given by $\left\vert \chi\right\rangle =\alpha\left\vert
\nearrow\right\rangle +\beta\left\vert \searrow\right\rangle $ with
$\alpha,\beta$ assumed to be real (a generalization for any point on the Bloch
sphere is possible but will not be needed here). Let us introduce the quantity
$|\alpha-\beta|^{2}$ that we may call the \textquotedblleft net diagonal
polarization\textquotedblright. Let $\left\vert 0_{q}\right\rangle $ denote
the ground state of the qubit pointer. The photon and qubit evolve from
$\left\vert \chi\right\rangle \left\vert 0_{q}\right\rangle $ to
\begin{align}
&  e^{-ig\sigma_{1}\sigma_{2}^{q}}\left\vert \chi\right\rangle \left\vert
0_{q}\right\rangle \\
&  =\left(  e^{-ig\sigma_{2}^{q}}\alpha\left\vert \nearrow\right\rangle
+e^{ig\sigma_{2}^{q}}\beta\left\vert \searrow\right\rangle \right)  \left\vert
0_{q}\right\rangle \\
&  =\alpha\left\vert \nearrow\right\rangle \left(  \cos g\left\vert
0_{q}\right\rangle +\sin g\left\vert 1_{q}\right\rangle \right)
+\beta\left\vert \searrow\right\rangle \left(  \cos g\left\vert 0_{q}
\right\rangle -\sin g\left\vert 1_{q}\right\rangle \right) \\
&  =\cos g\left(  \alpha\left\vert \nearrow\right\rangle +\beta\left\vert
\searrow\right\rangle \right)  \left\vert 0_{q}\right\rangle +\sin g\left(
\alpha\left\vert \nearrow\right\rangle -\beta\left\vert \searrow\right\rangle
\right)  \left\vert 1_{q}\right\rangle .
\end{align}
If the polarization is then measured in the linear basis and post-selected in
state $\left\vert H\right\rangle $, the qubit is left in state
\begin{equation}
\left(  \alpha+\beta\right)  \left\vert 0_{q}\right\rangle \cos g+\left(
\alpha-\beta\right)  \left\vert 1_{q}\right\rangle \sin g.
\end{equation}
For definiteness let us set $g=\pi/4.$ If $\left\vert \chi\right\rangle
=\left\vert H\right\rangle ,$ the net diagonal polarization is $0$ and the
qubit detector remains in its ground state. For $\left\vert \chi\right\rangle
=\left\vert V\right\rangle $ the detector is found with unit probability in
the excited state. \newline

Now in a Mach-Zehnder set-up, if we place a qubit pointer in state $\left\vert
0_{q}\right\rangle _{A}$ in arm A and apply $H_{int}=\gamma(t)\hat{X}%
_{A}\sigma_{2}^{q}$ the evolution of the joint system-pointer state can be
given as (keeping $g=\pi/4$)%
\begin{align}
&  e^{-ig\hat{X}_{A}\sigma_{2}^{q}}|A\rangle|H\rangle|0_{q}\rangle
_{A}+|B\rangle|{V}\rangle|0_{q}\rangle_{A}\\
&  =\frac{1}{\sqrt{2}}\left(  |A\rangle|H\rangle|0_{q}\rangle_{A}
+|A\rangle|V\rangle|1_{q}\rangle_{A}\right)  +|B\rangle|{V}\rangle
|0_{q}\rangle_{A}%
\end{align}
Postselecting to $\left\vert \phi\right\rangle $ [Eq. (\ref{post})] leaves the
pointer in the ground state $|0_{q}\rangle_{A}$ with unit probability.
Similarly, a qubit pointer in arm B coupled through $H_{int}=\gamma(t)\hat
{X}_{B}\sigma_{2}^{q}$ leads to the evolved state $\left(  |B\rangle
|V\rangle|0_{q}\rangle_{B}+|B\rangle|H\rangle|1_{q}\rangle_{B}\right)
/\sqrt{2}+|A\rangle|{H}\rangle|0_{q}\rangle_{B}$. The term $|B\rangle
|H\rangle|1_{q}\rangle_{B}$ is compatible with postselection, so the qubit
pointer can be found in the excited state $\left\vert 1_{q}\right\rangle _{B}$
with a non-zero probability.

\subsubsection{Disturbing joint measurement}

Let us finally consider the joint intermediate measurement of the spatial DOF
on arm A ($\hat{Y}_{A}$) and of the diagonal polarization DOF on arm B
($\hat{X}_{B})$ via coupling Hamiltonians $g\hat{Y}_{A}\hat{P}_{y}$ and
$g\hat{X}_{B}\hat{P}_{x}$ respectively, where $\hat{P}_{y}$ ($\hat{P}_{x})$
refers to the momentum of the pointer in arm A (arm B). Let $\left\vert
\xi_{y_{0}}\right\rangle $ and $\left\vert \xi_{x_{0}}\right\rangle $ denote
the initial states of the pointer in arm A and arm B respectively. The
evolution of the joint system-pointer state can be written as
\begin{align}
\left\vert \Psi(t)\right\rangle  &  =e^{-ig\hat{Y}_{A}\hat{P}_{y}/\hbar
}e^{-ig\hat{X}_{B}\hat{P}_{x}/\hbar}\left\vert \psi\right\rangle \left\vert
\xi_{y_{0}}\right\rangle \left\vert \xi_{x_{0}}\right\rangle \\
&  =\frac{1}{\sqrt{2}}\left\vert A\right\rangle \left\vert H\right\rangle
\left\vert \xi_{y_{0}+g}\right\rangle \left\vert \xi_{x_{0}}\right\rangle
+\frac{1}{2}\left\vert B\right\rangle \left\vert \xi_{y_{0}}\right\rangle
\left(  \left\vert \nearrow\right\rangle \left\vert \xi_{x_{0}+g}\right\rangle
-\left\vert \xi_{x_{0}-g}\right\rangle \left\vert \searrow\right\rangle
\right)  .
\end{align}
After post-selecting to state (\ref{post}) (and assuming the shifted pointer
states are orthogonal to their initial states) we will either find the pointer
coupled to the spatial DOF on arm A shifted by an amount $g$ along the $y$
axis and the pointer on arm B coupled to the diagonal polarization DOF
unshifted; or we will find that the pointer on arm $B$ has shifted by $\pm g$,
while the pointer on arm A remains unshifted.

Summing up the different cases we have considered concerning measurements at
an intermediate time inside the interferometer, we therefore see that:

\begin{itemize}

\item If only the spatial DOF is measured, the photon is always found on path
A, never on arm B.

\item If the diagonal polarization DOF is measured then it will be found in
arm B with a non-zero probability. If the diagonal polarization DOF is
coupled  to a qubit detector, such a detector will never be triggered in arm A.

\item If the spatial and the diagonal polarization DOFs are measured on paths
A\ and B respectively, the photon (as measured by its spatial degree of
freedom) is found on path A only half of the times.\ The other half, the
photon's diagonal polarization measurement is successful on path B, with the
pointer on arm B shifting in either the positive or negative directions.
\end{itemize}

Of course, these statements refer to three different setups. Quantum
measurements disturb the entire setup comprising the photon and the
measurement devices. It is not possible (except in a counterfactual sense) to
make any statement on a property of the photon that was not measured.\ Indeed,
these statements do not hold for the same photon, but for distinct particles
moving inside an inteferometer with distinct experimental arrangements.

\subsection{Weak measurements and weak values in the Mach-Zehnder
interferometer}

\subsubsection{Weak values}

We consider the same Mach-Zehnder interferometer and pre and post selected
states $\left\vert \psi\right\rangle $ and $\left\vert \phi\right\rangle $
given above [see Eqs. (\ref{pre}) and (\ref{post})]. An intermediate weak
measurement involves the same type of von Neumann interaction that was given
above for strong measurements [see Eq.(\ref{ecu0})] in a pre/post-selected
context.\ More precisely, let $\hat{S}$ denote the system obervable and assume
a system-pointer interaction $H_{int}=\gamma(t)\hat{S}\hat{P}$ coupling
$\hat{S}$ to $\hat{P}$ (the momentum of the pointer). The ensuing unitary
evolution operator $\exp\left(  -i\int H(t^{\prime})dt^{\prime}/\hbar\right)
$ brings the inital state $\left\vert \Psi(t_{i})\right\rangle =\left\vert
\psi\right\rangle \left\vert \xi_{x_{0}}\right\rangle $ to%
\begin{align}
\left\vert \Psi(t)\right\rangle  &  =e^{-ig\hat{S}\hat{P}/\hbar}\left\vert
\psi\right\rangle \left\vert \xi_{x_{0}}\right\rangle \\
&  \simeq\left(  I-ig\hat{S}\hat{P}/\hbar\right)  \left\vert \psi\right\rangle
\left\vert \xi_{x_{0}}\right\rangle
\end{align}
where $\left\vert \xi_{x_{0}}\right\rangle $ is the unshifted initial pointer
state and $g=\int\gamma(t)dt$ is now considered to be small so that the first
order asymptotic expansion holds. Actually more stringent conditions need to
hold \footnote{See e.g. I. M. Duck, P. M. Stevenson, and E. C. G. Sudarshan,
Phys. Rev. D 40, 2112, 1989.}, involving a sufficient width of the pointer
state and the transition amplitudes of $\hat{S}$. Successful post-selection
implies the projector $\Pi_{\phi}=\left\vert \phi\right\rangle \left\langle
\phi\right\vert $ is applied to the system state. The system-pointer state
after post-selection becomes%
\begin{align}
&  \left\vert \phi\right\rangle \left\langle \phi\right\vert \left.
\psi\right\rangle \exp\left(  -igS^{w}\hat{P}/\hbar\right)  \left\vert
\xi_{x_{0}}\right\rangle ,\label{wvpre}\\
=  &  \nu\left\langle \phi\right\vert \left.  \psi\right\rangle \left\vert
\phi\right\rangle \left\vert \xi_{x_{0}+g\operatorname{Re}(S^{w}%
)}\right\rangle , \label{wve}%
\end{align}
where we have re-exponentiated the term%
\begin{equation}
S^{w}=\frac{\left\langle \phi\right\vert \hat{S}\left\vert \psi\right\rangle
}{\left\langle \phi\right\vert \left.  \psi\right\rangle } \label{wvd}%
\end{equation}
known as the \emph{weak value } \footnote{The weak value was  introduced in Y.
Aharonov, D. Z. Albert, and L. Vaidman, Phys. Rev. Lett. 60,  1351 (1988).} of
$\hat{S}$ given the pre and post-selected states $\left\vert \psi\right\rangle
$ and $\left\vert \phi\right\rangle .$ $\nu$ is a normalization constant.

In general $S^{w}$ is a complex number. Since $\exp(-ia\hat{P})$ with $a$ real
is a translation operator, the pointer is shifted by $g\operatorname{Re}%
(S^{w}).$ The shift is small (relative to the pointer width) implying enough
statistics must be gathered in order to observe the shift position with low
uncertainty. Note this shift is the result of the interaction of the pointer
momentum with the system observable, and of the post-selection. When
$\operatorname{Re}(S^{w})=0$, the pointer does not shift despite the coupling
interaction. Operationally the property represented by $\hat{S}$ is not
recorded by the pointer when the system is post-selected. This can be
interpreted \footnote{See A. Matzkin, Found. Phys. 49, 298?316 (2019) and
Refs. therein.} as the absence, at the location of the pointer, of the
property corresponding to the coupled system observable. The imaginary part
$\operatorname{Im}S^{w}$ is interesting in its own right \footnote{See J.
Dressel and A. N. Jordan, Phys. Rev. A 85, 012107 (2012).}, but in the present
context it plays no role and we simply include it in the normalization
constant $\nu$.

\subsubsection{Joint weak values of the spatial and diagonal polarization
degrees of freedom}

Given our pre and post-selected states, it is straightforward to apply Eq.
(\ref{wvd}) to $\hat{Y}_{A}$. This yields the weak value of the spatial
projector on path A,%
\begin{align}
Y_{A}^{w}  &  =\frac{\left\langle \phi\right\vert \hat{Y}_{A}\left\vert
\psi\right\rangle }{\left\langle \phi\right\vert \left.  \psi\right\rangle }\\
&  =\left\langle A\right\vert \left\langle H\right\vert \left.  A\right\rangle
\left\vert H\right\rangle \\
&  =1.
\end{align}
$Y_{B}^{w}$ is not independent from $Y_{A}^{w}$, since by Eq. (\ref{proj-c})
we must have for any pre and post-selected states%
\begin{equation}
Y_{A}^{w}+Y_{B}^{w}=\frac{\left\langle \phi\right\vert \left.  \psi
\right\rangle }{\left\langle \phi\right\vert \left.  \psi\right\rangle }=1.
\end{equation}
Therefore $Y_{B}^{w}=0$ holds algebraically by virtue of Eq. (\ref{proj-c}).

For the diagonal polarization $\hat{X}_{i}=\left\vert i\right\rangle
\left\langle i\right\vert \otimes\sigma_{1}$, we have on path B
\begin{align}
X_{B}^{w}  &  =\frac{\left\langle \phi\right\vert \hat{X}_{B}\left\vert
\psi\right\rangle }{\left\langle \phi\right\vert \left.  \psi\right\rangle }\\
&  =\left\langle B\right\vert \left\langle H\right\vert \sigma_{1}\left\vert
B\right\rangle \left\vert V\right\rangle \\
&  =1.
\end{align}
$X_{A}^{w}$ is not independent from $X_{B}^{w}$, since%
\begin{align}
X_{A}^{w}+X_{B}^{w}  &  =\frac{\left\langle \phi\right\vert \sigma
_{1}\left\vert \psi\right\rangle }{\left\langle \phi\right\vert \left.
\psi\right\rangle }\\
&  =1,
\end{align}
where the last equality is state-dependent and holds for our specific choices
of pre and post-selected states, leading to $X_{A}^{w}=0$.

The upshot is that
\begin{align}
Y_{A}^{w}  &  =1\Rightarrow Y_{B}^{w}=1-Y_{A}^{w}=0\\
X_{B}^{w}  &  =1\Rightarrow X_{A}^{w}=1-X_{B}^{w}=0.
\end{align}
It is therefore necessary to measure jointly only the two weak values
$Y_{A}^{w}$ and $X_{B}^{w}$ on the same photon in order to observe
conclusively the separation between the spatial degree of freedom on path A,
and the diagonal polarization degree of freedom on path B.

\end{document}